# Pseudospin induced chirality with Staggered Optical Graphene

Jianlong Liu[1,2], Weimin Ye[1,3*], Shuang Zhang[1†]

1. School of Physics & Astronomy, University of Birmingham Birmingham, B15 2TT , UK

2. Department of Physics, Harbin Institute of Technology, Harbin, 150001, China

3. College of Optoelectronic Science and Engineering, National University of Defense Technology, Changsha, 410073, China

**Abstract:** Pseudospin plays a very important role in understanding various interesting physical phenomena associated with 2D materials such as graphene. It has been proposed that pseudospin is directly related to angular momentum, and it was recently experimentally demonstrated that orbit angular momentum is an intrinsic property of pseudospin in a photonic honeycomb lattice. However, in photonics, the interaction between spin and pseudospin for light has never been investigated. In this Letter, we propose that, in an optical analogue of staggered graphene, i.e. a photonic honeycomb lattice waveguide with in-plane inversion symmetry breaking, the pseudospin mode can strongly couple to the spin of an optical beam incident along certain directions. The spin-pseudospin coupling, caused by the spin-orbit conversion in the scattering process, induces a strong optical chiral effect for the

---

\* wmye72@126.com

† s.zhang@bham.ac.uk

**transmitted optical beam. Spin-pseudospin coupling of light opens door to the design of pseudospin-mediated spin or valley selective photonic devices.**

# INTRODUCTION

Graphene and Graphene-like two-dimensional (2D) materials have attracted great research interest in recent years [1-5]. Particular attention is paid to their unique electronic band structures which exhibit linear dispersion near the Dirac points at the corners of the Brillouin zone. This unusual band feature gives rise to many interesting electron transport properties, including quantum hall effect [1, 2], Zitterbewegung [3], and Klein paradox [4]. For single layer graphene deposited on substrate such as BN and SiC, the inversion symmetry between the two sublattices is broken. This symmetry breaking opens up gaps at the Dirac points, leading to the interesting optical selection rule at different valleys [6-10]. Specifically it was found that optical transitions at the two valleys are excited by light of different circular polarizations, or spins [11-13]. Inversion symmetry breaking also leads to lift of degeneracy between the two sublattice pseudospins. In contrast, for an ideal graphene without staggering potential, pseudospin is considered to be unmeasurable and cannot interact with any magnetic field even though it is suggested to exhibit real orbital angular momentum (OAM)[14, 15].

Recently, two dimensional photonic crystals with hexagonal lattice have provided a successful platform for demonstrating optical analogues of some of the interesting electronic properties in graphene [16]. Recent advances in this field include demonstration of optical spin Hall effect [17], discovery of unconventional edge states [18], demonstration of photonic Floquet topological insulators [19], and a number of other interesting optical phenomena based on metasurfaces, photonic and plasmonic crystals [20-22]. With artificial optical Graphene, it has been shown that pseudospin is indeed a measurable physical quantity. In particular, multiple-beam interference carefully aligned to the honeycomb lattice has been employed to excite the pseudospin modes, which were shown to exhibit OAM [23]. In this paper, we show that, in a staggered optical graphene (SOG) in which the inversion symmetry is broken between the two sublattices, pseudospin optical modes can be directly excited by a single circularly polarized beam with the aid of spin-pseudospin coupling, leading to strong optical activity for an incident wave with in-plane wave-vector matching the Dirac points of the lattice. Therefore, SOG not only represents a facile, lossless approach for achieving strong optical chirality which normally entails complex 3D metallic chiral structures, but also provides a platform for investigating extraordinary phenomena associated with the pseudospin state of light in a honecomb lattice.

In previous studies, artificial photonic Graphenes have been realized by using coupled optical resonators or waveguides, where each resonator or waveguide serves as the optical analogue of a carbon atom in graphene. Consequently, the coupling between the adjacent resonators or waveguides can be treated as photon hopping in a similar manner as the electron hopping in their electronic counterpart. Under these conditions, tight-binding approximation and Hamiltonian approach can be conveniently adopted [14, 17, 23, 24]. Here, without resorting to the tight-binding conditions, we first show that the OAM is an intrinsic property of pseudospin in a photonic honeycomb lattice. The derivation is purely based on symmetry operations, with details given below.

## MATERIALS AND METHODS

The schematic of a 2D honeycomb lattice for investigation of the spin-pseudospin coupling is shown in Figure 1a. The primitive lattice vectors of the honeycomb lattice are $a_1 = a\hat{x}$ and $a_2 = (\hat{x} + \sqrt{3}\hat{y})a/2$. We consider the eigenmode of the 2D honeycomb lattice with in-plane vector $K = -4\pi/3a\hat{x}$ (Figure 1b). It is obvious that the in-plane vector $K$ is unchanged under the rotation operation $C(\hat{z}, 2\pi/3) = C_3$ with respect to the center of the hexagon cell $A_1B_1A_2B_2A_3B_3$. The action $C_3$ rotational operator on the eigenmode at $K$ point can be written as [25]

$$C_3 \circ \psi_K(X) = R_{C_3} \psi_K\left(R_{C_3}^{-1} X\right) \tag{1}$$

where,

$$R_{C_3} = \begin{pmatrix} \cos\frac{2\pi}{3} & -\sin\frac{2\pi}{3} & 0 \\ \sin\frac{2\pi}{3} & \cos\frac{2\pi}{3} & 0 \\ 0 & 0 & 1 \end{pmatrix}, \quad R_{C_3}^{-1} X = \begin{pmatrix} \cos\frac{2\pi}{3} & \sin\frac{2\pi}{3} & 0 \\ -\sin\frac{2\pi}{3} & \cos\frac{2\pi}{3} & 0 \\ 0 & 0 & 1 \end{pmatrix} \begin{bmatrix} x \\ y \\ z \end{bmatrix}$$

From Figure 1a we can write,

$$R_{C_3}^{-1} X_{A_1} = X_{A_3} = X_{A_1} + a_1; \quad R_{C_3}^{-1} X_{B_1} = X_{B_3} = X_{B_1} + a_2$$

where $X_{A_1}$ and $X_{B_1}$ are the positions of element $A_1$ and $B_1$ in a hexagon cell. So, Equation (1) can be rewritten as

$$C_3 \circ \psi_K(X) = R_{C_3} \psi_K\left(R_{C_3}^{-1}(X - X_{A_1}) + X_{A_1} + a_1\right) = R_{C_3} \psi_K\left(R_{C_3}^{-1}(X - X_{B_1}) + X_{B_1} + a_2\right) \quad (2)$$

As the lattice has translational invariance, the eigenmode of the 2D honeycomb lattice satisfies

$$\psi_K(X + ma_1 + na_2) = e^{iK \cdot (ma_1 + na_2)} \psi_K(X).$$

Since in-plane vector $K$ is unchanged under the rotation operation $C_3$, it can be deduced from Equation (2) that

$$C_3 \circ \psi_K(X) = e^{iK \cdot a_1} R_{C_3} \psi_K\left(R_{C_3}^{-1}(X - X_{A_1}) + X_{A_1}\right) = e^{iK \cdot a_2} R_{C_3} \psi_K\left(R_{C_3}^{-1}(X - X_{B_1}) + X_{B_1}\right)$$

which can be rewritten as,

$$C_3 \circ \psi_K(X) = e^{i\frac{2\pi}{3}} R_{C_3} \psi_K\left(R_{C_3}^{-1}(X - X_{A_1}) + X_{A_1}\right) = e^{-i\frac{2\pi}{3}} R_{C_3} \psi_K\left(R_{C_3}^{-1}(X - X_{B_1}) + X_{B_1}\right) \quad (3)$$

Based on Equation (1), Equation (3) can be rewritten by using the rotation operation $C(\hat{z}, 2\pi/3)$ with respect to the centers of element $A_1$ and $B_1$ as

$$C_3 \circ \psi_K(X) = e^{i\frac{2\pi}{3}} C_3(X_{A_1}) \circ \psi_K(X) = e^{-i\frac{2\pi}{3}} C_3(X_{B_1}) \circ \psi_K(X) \quad (4)$$

For the honeycomb structures with $C_3$ rotational symmetry, the eigenmode at $K$ point satisfies

$$C_3 \circ \psi_K(X) = e^{i\frac{2q\pi}{3}} \psi_K(X), \quad q = -1, 0, 1 \quad (5)$$

The OAM of an optical mode is directly related to its angular phase distribution surrounding a certain point. Alternatively, it is manifested as the phase acquired by the mode when it is rotated by certain angles (depending on the rotational symmetry of the system). Eq. (5) shows that rotating the eigenmode around the center of the hexagon by 120° introduces a phase terms of -2π/3, 0 or 2π/3 for q = -1, 0 or +1, respectively, corresponding to an OAM of 1, 0 or -1. Combing Equation (4) and (5), it can be deduced that

$$C_3(X_{A_1}) \circ \psi_K(X) = e^{i\frac{2(q-1)\pi}{3}} \psi_K(X); \quad C_3(X_{B_1}) \circ \psi_K(X) = e^{i\frac{2(q+1)\pi}{3}} \psi_K(X) \tag{6}$$

According to Equation (6), we can deduce that eigenmode with $q = -1$ has an OAM of $-1$ at sublattice A and 0 at sublattice B, eigenmode with $q = 0$ has an OAM of 1 at sublattice A and $-1$ at sublattice B, and eigenmode with $q = 1$ has an OAM of 0 at sublattice A and 1 at sublattice B. Due to the high spatial symmetry of $q = 0$ mode, it can hardly be excited by an incident plane wave. We therefore focus on the two modes with $q = \pm 1$. Similar to Graphene, the eigenmodes of $q = 1$ and $-1$ can be endowed with pseudospin $|+\rangle$ and $|-\rangle$ which possess OAM of (0, 1) and ($-1$, 0) at sublattice (A, B), respectively. Due to the spin-orbit conversion of light during the scattering process, the eigen-modes with different pseudospins can be selectively excited by external circularly polarized light of different handednesses.

## RESULTS AND DISCUSSION

To confirm the above theoretical analysis, we numerically study a realistic SOG based on photonic crystal slab with thickness $h$. The schematic illustration is depicted in Figure 1(c). The photonic lattice consists of two sets of circular apertures with diameter $d_A$ and $d_B$ as shown in Figure 1(c). In the simulation, a dielectric slab with refractive index $n = 3$ (corresponding to AlAs at wavelength around 800 nm) is used and other geometric parameters are set as $h = 0.2a$ and $d_A = 0.30a$, $d_B = 0.28a$. The band diagram for this structure is plotted in Fig. 2a. The dispersion relation of the eigen-modes near one **K** point $(-4\pi/3a, 0)$ is calculated using commercial FEM software COMSOL and plotted in Figure 2b. As shown in the figure, the eigen-frequencies ($\omega_0' = 1.048 \times 2\pi c/a$ and $\omega_0'' = 1.052 \times 2\pi c/a$) at **K** point lie inside the light cone, which means that the eigen-modes of the SOG can couple with external light fields directly. Figure 2c and 2d show the field distributions ($|H_z|$) at the symmetric plane in $z$ direction for the two eigenmodes with frequency $\omega_0'$ and $\omega_0''$, respectively. As expected, the field distributions show three fold rotational symmetry. The phase patterns (arg($H_z$)) of the modes obtained from the simulations are shown in Figure 2f and 2g, respectively. In Figure 2f, it is clear that there are phase vortices only at the lattice B accompanied with an opposite phase vortex ($q = 1$) at the symmetric center. While for the eigen-mode with $q = -1$ (Figure 2g), the opposite phase vortices are located only at the lattice A. In addition, we also calculate the amplitude and phase pattern of an eigen-mode with $q = 0$, which are shown in Figure 2e and 2h, respectively. The eigen-frequency is $\omega_0''' = 1.110 \times 2\pi c/a$, which is relatively far from the resonance frequencies of the other two modes. There exist opposite phase vortices at lattice A and B, while no phase vortex is present at the symmetric center (corresponding to $q = 0$). The simulation results are consistent with the theoretical analysis.

Here, we propose to use a circularly polarized optical beam to selectively excite the pseudospins in SOG. The spin angular momentum (SAM) of the incident light can be converted to OAM by light scattering upon subwavelength objects [26-28]. We use a numerical simulation to check the spin-orbit AM conversion. In our setting, a dielectric slab with a single air hole is illuminated by right/left circular polarized light (RCP/LCP), respectively. The diameter of the hole is 0.28$a$. Other parameters are kept the same as those in the mode calculation in previous sections. The frequency of the incident light is chosen as $1.048 \times 2\pi c/a$ which is the same as the eigen-frequency $\omega_0'$ in the mode calculation. The incidence is inclined with $k$ vector $(-4\pi/3a, 0)$. The simulation results including the amplitude ($|H_z|$) and phase ($\arg(H_z)$) distributions at the symmetric plane in $z$ direction are shown in Figure 3(a)-3(d), respectively. For an incident plane wave with circular polarization at normal incidence onto a lossless circular object, there is no transfer of the angular momentum between light and the object; the angular momentum must be conserved. Consequently, the wave scattered into the guided mode has an exact orbit angular momentum of 1 or -1 since the guided mode is linearly polarized and does not carry spin angular momentum. In our case, however, the incidence is inclined (almost 40°), the spin of light projected onto the plane is less than 1. This leads to nonuniform angular distribution of the field amplitude around the aperture, as shown in Figure 3a and 3b. Nonetheless, the phase distributions exhibit obvious opposite vorticities centered at the symmetric center of the air hole (Figure 3c and 3d). Such spin-orbit AM conversion provides a bridge linking the spin of light and the pseudospins of the SOG.

In order to demonstrate the spin-pseudospin coupling and the selective excitation of pseudospins, simulations are carried out to calculate the transmission spectrum of SOG. The incident circularly polarized light with fixed in-plane wave vector $K\left(-4\pi/3a,0\right)$ illuminates the photonic slab as shown in Figure 1(c). All geometric parameters are the same as those in Figure 2b. The zero order transmittance is measured and presented in Figure 4a. There exists significant difference in transmission near the two eigen-frequencies between the RCP and LCP incidences, confirming the presence of very strong optical chirality. The key reason for such transmission difference is the broken inversion symmetry of the SOG structure. As the inversion symmetry is broken, the bandgap at Dirac point is open and the two pseudospin states possess different eigen-frequencies. When the incident RCP and LCP light couples with the pseudospins, respectively, the resonance peaks in the transmission spectra between the two circular polarizations are consequently distinct from each other. In our setting, the photonic crystal slab is not an ideal 2D sheet because of the finite thickness in $z$ direction. An analysis based on Fano line shapes, describing the interference between the scattered fields from the A B sublattices and the plane slab [29], is carried out. The fitting curve is expressed as

$$T(\omega) = 1 - C_1 \frac{\left(\alpha\Gamma/2 + \omega - \omega_0\right)^2}{\left(\Gamma/2\right)^2 + \left(\omega - \omega_0\right)^2} - C_2\left(\omega - \omega_d\right) \qquad (7)$$

where $\omega_0$ is the resonant frequency, $\Gamma$ is the linewidth (FWHM), $\alpha$ is the Breit-Wigner-Fano coupling coefficient [30], $C_1$, $C_2$ and $\omega_d$ are three coefficients to be fitted. The third term on the RHS of Equation (7) stands for the contribution of the direct transmission of a homogeneous slab with an effective refractive index. This term represents the background of the spectrum in Figure 4a. Because the frequency band we are concerned is very narrow, the background spectrum can assume a linear form fitted by $C_2$ and $\omega_d$ within this narrow range. The Fano-lineshape fitted curves are shown in Figure 4a along with the simulation data. The curves all agree well with the simulation data. The fitted resonant frequencies are in good accordance with the simulated eigen-frequencies in the band calculation in Figure 2b.

To verify the contribution of the spin-pseudospin coupling to the transmission spectrum, we examine the field distributions when the crystal is illuminated by RCP and LCP light. Same as in Figure 2c, 2d, 2f and 2g, we extract the amplitude and phase of the magnetic field at the symmetric plane in $z$ direction. The frequencies of incident light are fixed at the eigen-frequencies at K point. The results are all depicted in Figure 4(b)-4(e). The amplitude and phase pattern show good agreement with those of the eigen-modes depicted in Figure 2. It is thus confirmed that incident circularly polarized beam can excite the pseudospins with matched handedness.

It is worth mentioning that in our setting we use a free standing 2D photonic crystal where no substrate is added. Free-standing 2D photonic crystal membrane based on silicon or other semiconductor materials have been demonstrated by many groups [31-33]. Therefore the symmetric configuration investigated in our paper can be fabricated without posing significant technical challenges. On the other hand, the presence of a substrate does affect the performance. If the substrate (e.g. Glass) is added, the Q values of the leaky modes would decrease and consequently the contrast between the RCP and LCP transmission is expected to be reduced.

**CONCLUSIONS**

In summary, we have studied the spin-pseudospin coupling in a SOG structure. We predict and numerically demonstrate that the two pseudospin states at Dirac point of SOG can be directly excited by external circularly polarized light with opposite handedness. Consequently, we show that the transmission spectrum of the lattice exhibit strong chirality which arises from the coupling between spin and the intrinsic handedness of the pseudospin. Due to the inversion symmetry of the reciprocal space, the spin-orbit interaction from pseudospin leads to the coupling of spin and valley degrees of freedom, which makes it possible to selectively choose the spin of the incident light at different valleys. As this spin-valley coupling occurs inside the light cone, our work may also provide opportunities to construct valley-dependent circularly polarized light emitters or generators.

## ACKNOWLEDGEMENTS

This work was supported by EPSRC (Grant No. EP/J018473/1), National Nature Science Foundation of China (Grant Nos. 61307072, 11374367, 61328503), and Leverhulme Trust (Grant No. RPG-2012-674)

FIGURE CAPTURE LIST

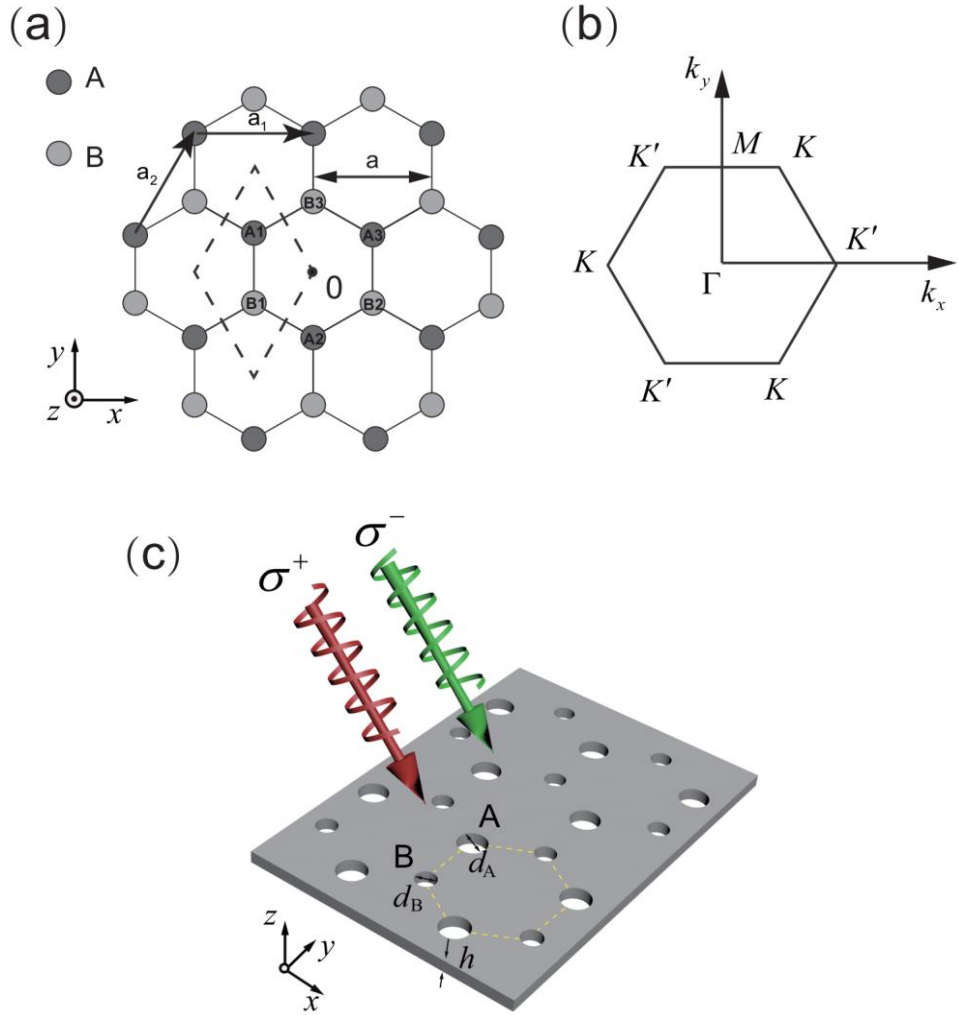

Figure 1. Illustration of the honeycomb lattice. (a) A honeycomb lattice with lattice constant $a$ and lattice vectors $\boldsymbol{a}_1 = a\hat{x}$ and $\boldsymbol{a}_2 = \frac{a}{2}(\hat{x}+\sqrt{3}\hat{y})$. (b) The first Brillouin zone in reciprocal space with the positions of $K$ and $K'$ indicated. (c) Schematic illustration of an artificial SOG based on a 2D photonic crystal slab. The slab is illuminated by a plane wave with circular polarizations.

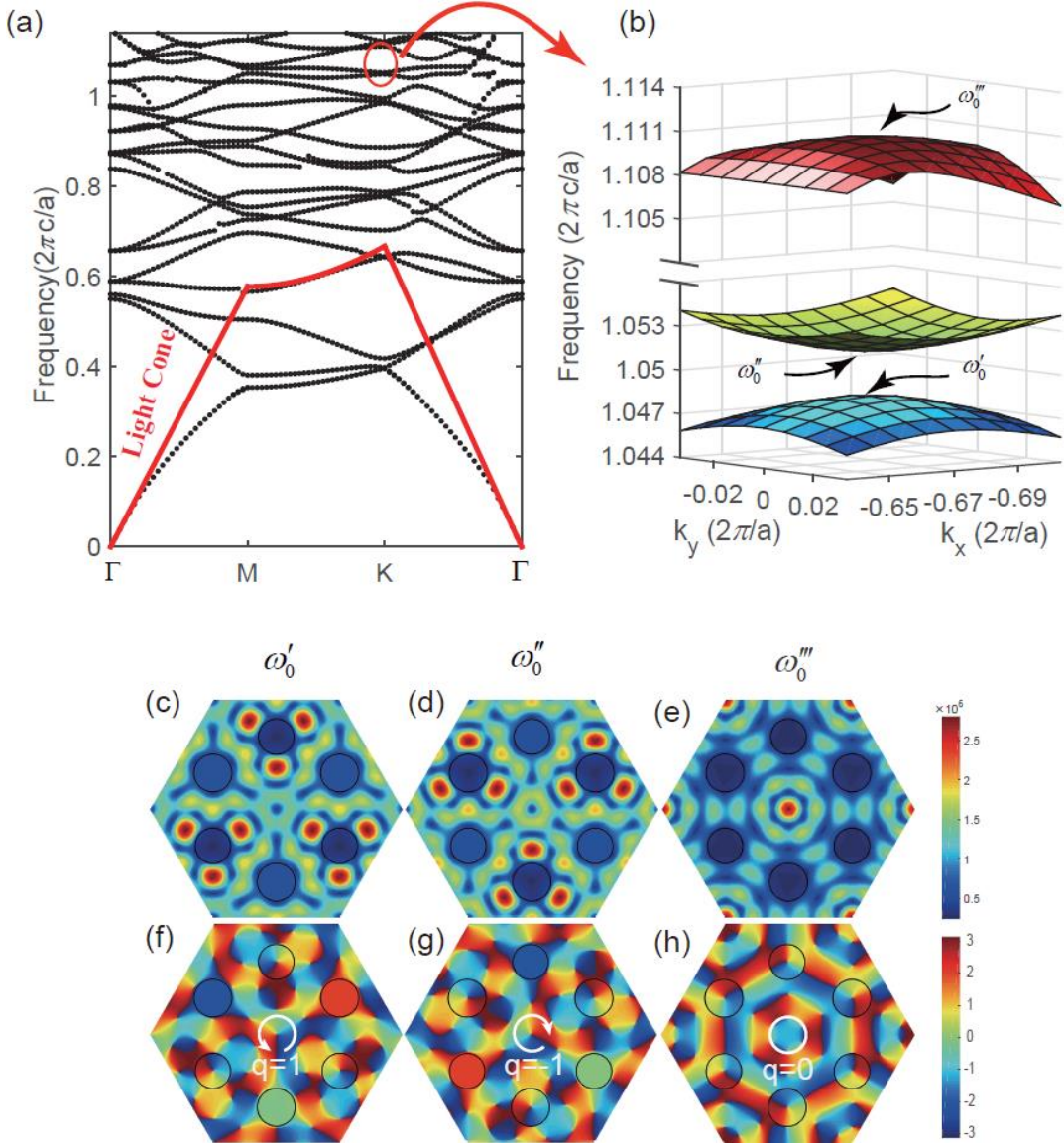

Figure 2. Eigen-modes supported by the 2D photonic lattice. (a) Band structure for the even modes in the SOG. (b) Dispersion relation in the vicinity of K point $(-4\pi/3a, 0)$. (c-e) Amplitude ($|H_z|$) distributions at the symmetric plane in $z$ direction for the eigen-modes at $K$ point for: $q = 1$ ($\omega_0' =$ 1.048×$2\pi c/a$), $q = -1$ ($\omega_0'' =$1.052×$2\pi c/a$), and $q = 0$ ($\omega_0''' =$ 1.110×$2\pi c/a$).   Note that the q = 0 mode is out the frequency range in (b). (f-h) phase distributions (arg($H_z$)) of the three modes corresponding to $q = +1$, -1 and 0.

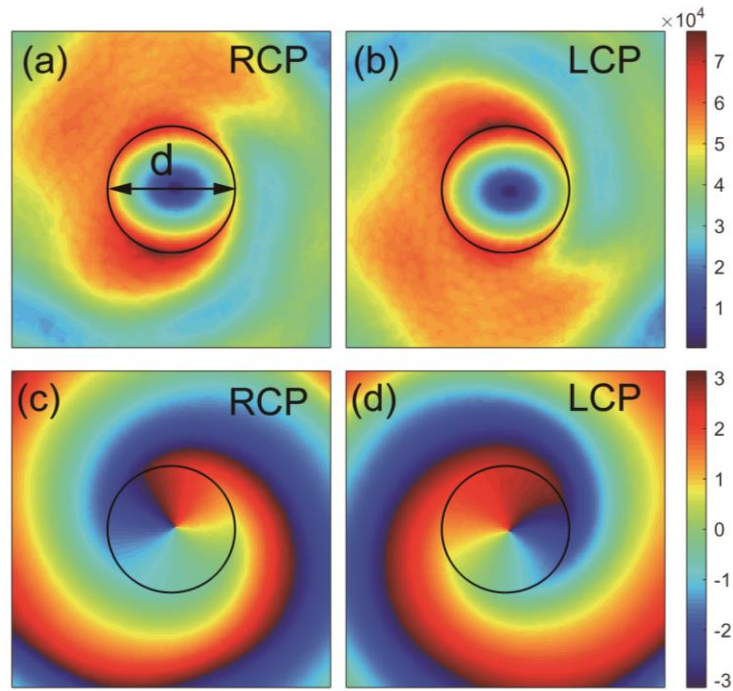

Figure 3. Conversion of SAM of incident light into OAM of the guided mode through scattering by an individual circular aperture. (a, b) Simulated field amplitude ($|H_z|$) distribution produced by an incident RCP (a) and LCP (b) beam. (c, d) The phase ($\arg(H_z)$) distribution produced by RCP (c) and LCP (d) incident wave. In the simulation, the diameter of the circular aperture $d = 0.28a$.

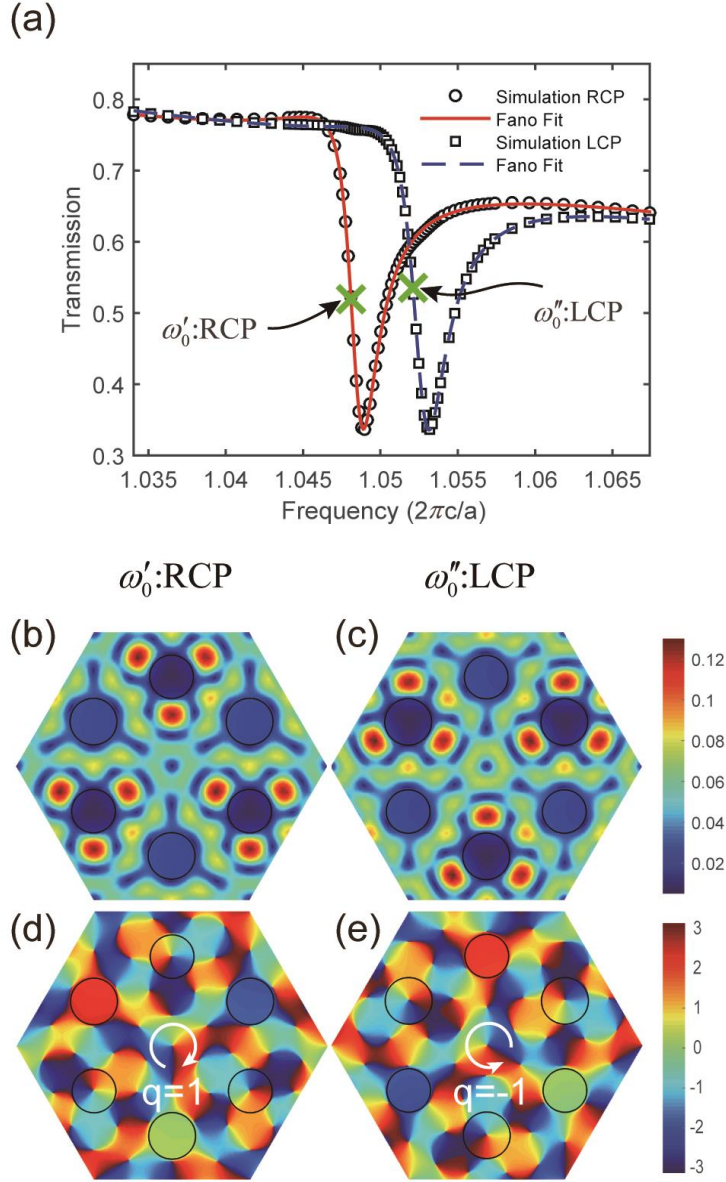

Figure 4. Chiral response of the 2D photonic crystal through spin-pseudospin coupling. (a) Simulated and Fano fitted spectra of zero order transmission for the SOG. The structure is illuminated by RCP and LCP light beam with in-plane wave vector along K point. The fitting parameters are $C_1 = 0.049$, $\alpha = 2.77$, $\Gamma = 2.2 \times 10^{-3}$, $\omega_0=1.048$, $C_2=3.04$, $\omega_d=0.971$ for RCP and $C_1=0.041$, $\alpha=2.98$, $\Gamma=2.6\times10^{-3}$, $\omega_0=1.053$, $C_2=3.39$, $\omega_d=0.978$ for LCP incidence. The eigenfrequencies of the pseudospin modes are indicated by the cross symbols. (b, c) Amplitude ($|H_z|$) at the symmetric plane of the slab in $z$ direction for RCP (b) and LCP (c) incidence at their respective eigenfrequencies. (d, e) The corresponding phase distribution of $H_z$.